\documentclass[conference]{IEEEtran}
\IEEEoverridecommandlockouts
\usepackage{cite}
\usepackage{amsmath,amssymb,amsfonts}
\usepackage{algorithmic}
\usepackage{graphicx}
\usepackage{textcomp}
\usepackage{xcolor}
\def\BibTeX{{\rm B\kern-.05em{\sc i\kern-.025em b}\kern-.08em
    T\kern-.1667em\lower.7ex\hbox{E}\kern-.125emX}}
\begin{document}

\title{Joint Modeling of Longitudinal and Survival Data with Censored Single-index Varying Coefficient Models\\
}

\author{\IEEEauthorblockN{Jizi Shangguan}
\IEEEauthorblockA{\textit{} \\
\textit{George Washington University}\\
Washington DC \\
shangguanjizi@gwmail.gwu.edu}\\
}

\maketitle

\begin{abstract}
In medical and biological research, longitudinal data and survival data types are commonly seen. 
Traditional statistical models mostly consider to deal with either of the data types, such as linear mixed models for longitudinal data, and the Cox models for survival data, while they do not adjust the association between these two different data types. It is desirable to have a joint modeling approach which accomadates both data types and the dependency between them. In this paper, we extend traditional single-index models to a new joint modeling approach, by replacing the single-index component to a varying coefficient component to deal with longitudinal outcomes, and accomadate the random censoring problem in survival analysis by nonparametric synthetic data regression for the link function. Numerical experiments are conducted to evaluate the finite sample performance. 
\end{abstract}


\section{Introduction }\label{sec:tobit-model)}

In many medical studies, repeated measurements or longitudinal data are frequently observed, such as patient health outcomes data or biomarker data at sequential visits for each patient \cite{ORKIN2020e389,huang2019decomposition,erlandson2021weight,acosta2021three}. In addition, survival outcomes are usually used as endpoints for clinical evaluations, such as time to disease progression, in particular in cancer trials\cite{huang2019bi,huang2020high,castagna2020548}. Different statistical methods have been proposed to deal with longitudinal data or survival data, respectively. For example, linear mixed models can be used to deal with repeated measurements, and Cox proportional hazard models could be used to deal with survival data. While these traditional models do not consider possible dependency between the two types of data, especially considering the fact that the two different types of data often appear simultaneously, for example, in cancer clinical trials, patient data are collected longitudinally, and progression free survival is often the clinical endpoints\cite{ibrahim2010basic,huang2019estimation}. Thus, it is desirable to have statistical methods to accomadate both data types at the same time and account for the dependency, where it is also referred as a "joint modeling" approach.

In the literature of statistical modeling, single-index models have been extensively studied and discussed as an extension of linear regression models, which retains the interpretability of linear regression models, and enables more model flexibility by imposing a nonparametric link function\cite{huang2020estimation,huang2020bi,huang2017semi}. Given the popularity of single-index models, in this article, we extend traditional single-index models, and propose a censored single-index varying coefficient  model to accomadate for longitudinal and survival data. The longitudinal data type is justified by the varying coefficient component and the survival data type is addressed by the non-parametric link function with a nonparametric synthetic data regression method.

The paper is organized as follows, section II  presents the model, the intuition and estimation methods. Section III presents the numerical experiments. Section IV states conclusion and further interest.

\section{Model and Estimation procedure}

\subsection{Model}
Consider the following censored single-index varying coefficient model,
\begin{eqnarray}
\label{eq:11}
Y_i^{*} =m(\sum_{j=1}^{d}\beta_{j}(T_{i})X_{i,j}) +\epsilon_i,~i=1,\ldots,n,
\end{eqnarray}
where $Y_i^{*}$ represents the survival time, $X_i = (X_{i,1}, \ldots, X_{i,d})^{\top}$ is a $d$-dimensional covariate vector,  each $\beta_j(\cdot)$ is an unspecified smooth functions on $[0,1]$, $m(\cdot)$ is an unknown smooth function, and $\epsilon_i$'s are random errors. Let $C$ be the random censoring time variable associated with the response $Y^{*}$, and we assume that $C$ is independent of $(X,Y^{*})$.
Due to random censoring, we can only observe $(Y_i,\delta_i)$, where $Y_i=\min(Y_i^*,C_{i})$, $\delta_i=I(Y_i^*<C_{i})$. Instead of making parametric distributional assumptions such as normality, here we only assume that $\epsilon_i$'s are independently and identically distributed (i.i.d.) from an unknown distribution symmetric around $0$ with finite variance.
Furthermore, we assume that no intercept is included in the index function $X_i^{\top}\beta(T_{i})$, $||\beta(t)||=1$ for each $t$ and the first element of $\beta(\cdot)$ is positive, to ensure identifiability, where $\|\cdot\|$ denotes the $L_2$ norm. In addition, $X\in D_X \subset \mathbb{R}^{d}$ for some compact set $D_X$.

\subsection{Estimation of varying coefficient components}

Under model (\ref{eq:11}), the proposed estimation procedure for $\beta_j(\cdot)$ is inspired by a theorem below.

\noindent{\bf Assumption A.1} (i) The latent response $Y^*$ has first $\nu (\ge 3)$ absolute moments.
(ii) The density functions of $\epsilon_i$ and censoring variable $C_{i}$, denoted as $f_{\epsilon}(\cdot)$ and $f_{C}(\cdot)$, are continuously differentiable, and $f_{\epsilon}(\cdot)$ is symmetric around zero.

\noindent{\bf Theorem 1}
	Let $F_\epsilon(\cdot)$ be the distribution function of $\epsilon$. Under Assumption A.1, if $\lim_{\epsilon_{i}\to-\infty} \epsilon F(\epsilon)$=0,
	then 
\begin{eqnarray}
&&E(Y_i|X_{i}^{\top} \beta(T_{i}),T_{i}) \nonumber\\
&=& \int_{-\infty}^{+\infty} \Big[-\int_{-\infty}^{c-m(X_{i}^{\top} \beta(T_{i}))} F(\epsilon)d\epsilon \nonumber\\
&&+cF\Big\{c-m(X_{i}^{\top} \beta(T_{i}))\Big\}\Big] f_{C}(c) dc d\epsilon. 
\end{eqnarray}

Proof of Theorem 1 could be referred to proof of proposition 1 in \cite{huang2020high}.
Assumption A.1 and the assumption $\lim_{\epsilon\to-\infty} \epsilon F(\epsilon)$=0 are mild, and most commonly used distributions, like normal distribution, Student $t$ distribution, uniform distribution on a symmetric interval, satisfy these assumptions \cite{huang2020high}.
Theorem 1 implies that $E(Y_i|X_i^{\top}\beta(T_{i}))$ can be represented as a new model with the same varying coefficient functions $\beta_{j}(\cdot)'s$, but a new link function, i.e., $E(Y_{i}|X_{i}^{\top}\beta(T_{i})=u)=r(u)=w \circ m(u)$, where '$\circ$' means composition of two functions and 
$$w(t)=\int_{-\infty}^{+\infty} \Big[-\int_{-\infty}^{c-t} F(\epsilon)d\epsilon+c F\Big\{c-t\Big\}\Big] f_{C}(c) dc d\epsilon.$$

By Theorem 1, we give a new single-index varying coefficient for the observed response as
\begin{equation}
\label{eq:2}
Y_{i}=r(X_{i}^{\top}\beta(T_{i})) + \epsilon_{i}^{'},  i = 1,2,\ldots, n,
\end{equation}
where $\epsilon_{i}^{'}=\epsilon_{i}+(-Y^{*}_{i}+Y_{i})-r(X_{i}^{\top}\beta(T_{i}))+m(X_{i}^{\top}\beta(T_{i}))$, satisfying $E(\epsilon^{'}_{i}|X_i^{\top} \beta)=0$ given Assumption A.1.
By re-expressing our orginal model into (\ref{eq:2}), a general estimation method for single-index varying coefficient models can be applied to estimate $\beta_{j}(\cdot)'s$. We adopt the Least-Squares type method in \cite{kuruwita2011generalized} to estimate  $\beta_{j}(\cdot)$, denoted as $\hat{\beta_{j}}(\cdot)$. 

\subsection{Nonparametric synthetic data regression of $m(\cdot)$ }

With estimated $\hat{\beta_{j}}(\cdot)$ in hand, we now move to discuss the estimation of the unknown link function $m(\cdot)$.
Let $\tau_{i}=X_{i}^{T}\beta(T_{i})$.
To begin, we re-express model(\ref{eq:11}) as (\ref{eq:3}).  With estimates $\hat{\beta_{j}}(\cdot)$ in hand, we now estimate the unknown link function $m(\cdot)$.
To begin, we re-express model (\ref{eq:11}) as
\begin{equation}
\label{eq:3}
Y_i^*=m(\tau_{i})  + \epsilon_i,~i=1,\ldots,n,
\end{equation}
where $U_{i}=X_{i}^{\top}\hat{\beta}(T_{i})$.
Therefore, we have an one dimensional nonparametric randomly censored regression model with predictor variable $U_{i}$, 
traditional methods like \cite{fan1994censored,koul1981regression} can be used to estimate $m(\cdot)$. For computational and theoretical convenient, we adopt the method in \cite{koul1981regression} and give the following estimating procedure for $m(\cdot)$.
Let $G(t)=P(C \ge t)$, and define the synthetic transformation of response variable as 
$$T^{*}_{i}=\int_{0}^{\infty}(\frac{|Y_{i} \ge s|}{\hat{G}(s)}),$$
where $\hat{G}(t)$ is the Kaplan-Meier estimator of the survival function $G(t)$.
After transformation, usual kernel smoothing can be applied to observations $(X_{i},T^{*}_{i})$ to estimate the link function $m(\cdot)$. The Nadaraya-Watson kernel estimator of $m(x)$ based on the synthetic data $(X_{i},T^{*}_{i})$ is given by
$$\hat{m}(x)=n^{-1}\sum_{i=1}^{n}T^{*}_{i}K_{h}(x-X_{i})/f_{n}(x),$$
where $f_{n}(x)=n^{-1}\sum_{i=1}^{n}K_{h}(x-X_{i})$, the Rosenblatt-Parzen estimator of $f(x)$. And $K_{h}(x-X_{i})=h^{-d}K((x-X_{i})/h)$ , where $K(\cdot)$ is a $d$-variate kernel function. The bandwidth selection follows the same pattern as in \cite{kuruwita2011generalized}.


\section{Simulation study}

In this section, we investigate the finite sample performance of our proposed estimation by Monte Carlo simulations.
We generated $100$ random samples of size $n$ responses from model $(1)$ with two covariates. We took $X1,X2$ to be 
to be independent standard normal covariates. The effect modifier $U$ is taken to be a uniform random variable over $[0, 1]$ and $\epsilon_i \sim N(0,0.2^{2})$, independent of both $X$ and $U$. The data are generated from model(\ref{eq:11}):
$m(u)=u^2$ and $\beta_{1}(t)$,  $\beta_{2}(t)$ were taken to be $cos(t)$ and $sin(t)$. $n=500$.  Choose $c$ such that  the censoring level is $0.3$.
In order to assess the pointwise variability of the coefficient function estimators, we plotted the 5th and the 95th pointwise percentiles of the 100 estimates of our estimated coefficient functions. 
Figure 1 presents the point-wise median curve of the estimated functions $\hat{\beta}_{1}(\cdot)$ and $\hat{\beta}_{2}(\cdot)$ on the grid points uniformly spaced between $[0,1]$, point-wise 5\% quantile and 95 \% quantile of  $\hat{\beta}_{1}(\cdot)$ and $\hat{\beta}_{2}(\cdot)$. 
In general, the fitted values (solid lines) are close to the true values (dashed lines), and the confidence bands covers the true function except for some small region. 
And Figure 2 presents the point-wise median curve of the estimated function $\hat{m}(\cdot)$ n the grid points uniformly spaced between $[-0.5,0.5]$, with 5\% quantile and 95 \% quantile of  $\hat{m}(\cdot)$, it can be seen that it assembles a quadratic function.

\begin{figure}[pbth]
	\includegraphics[width=0.5\textwidth,height=0.3\textheight]{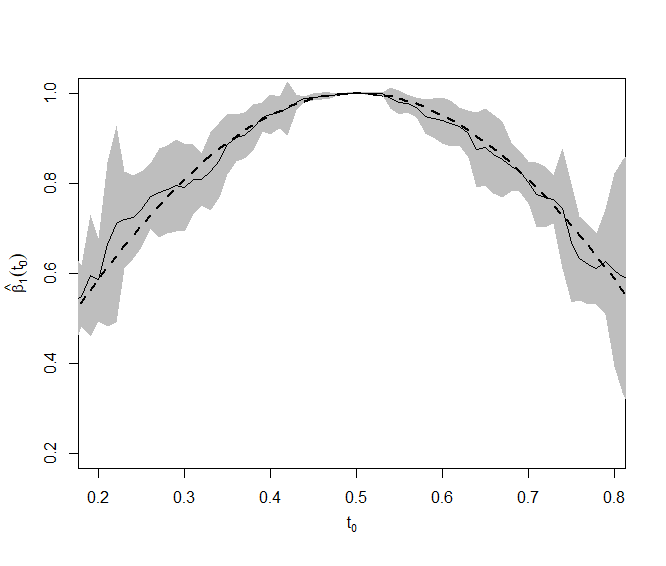}
	\includegraphics[width=0.5\textwidth,height=0.3\textheight]{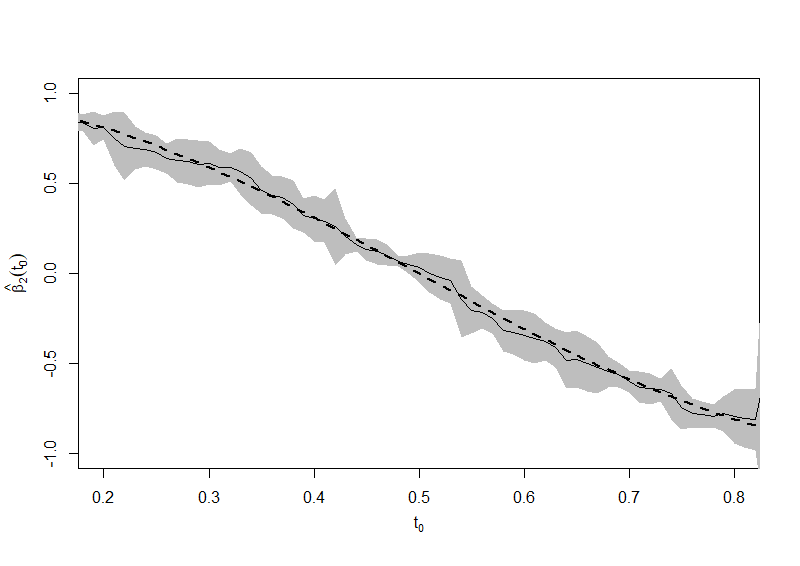}
	\caption{Varying Coefficient Components Estimation}
\end{figure}

\begin{center}
	\begin{figure}[pbth]
		\centerline{\includegraphics[width=0.5\textwidth,height=0.3\textheight]{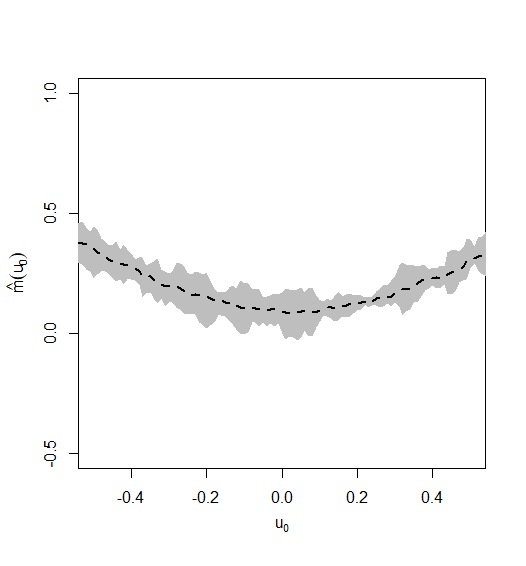}}
		\caption{Link Function Estimation}
	\end{figure}
\end{center}

\section{Conclusions}
In this research article, we proposed a new censored single-index varing coefficient model as a tool for joint modeling of longitudinal and survival data.Varying coefficient index adjusts for the longitudinal effect and the non-parametric link function accounts for random censoring problem. The intuition of the method was stated and numerical experiments were examined. In the future, research directions would be in asymptotic properties and real data examples.

\bibliographystyle{IEEEtran}
\bibliography{tobit_refall_new}{}

\end{document}